\documentclass[a4paper,11pt]{article}
\pdfoutput=1 

\usepackage{jheppub} 

\usepackage[T1]{fontenc} 

\usepackage{mathtools,amsthm,amsfonts} 
\usepackage{graphicx} 
\usepackage{fancyhdr,titlesec,microtype} 
\usepackage{mathrsfs}
\usepackage{url}
\usepackage{enumitem}
\usepackage{lineno}

\def\beq{\begin{equation}}
\def\eeq{\end{equation}}
\def\ba{\begin{array}}
\def\ea{\end{array}}
\def\bea{\begin{eqnarray}}
\def\eea{\end{eqnarray}}

\def\G{\Gamma}
\def\S{{\bf S}}

\def\CA{{\cal A}}
\def\CB{{\cal B}}
\def\CC{{\cal C}}

\def\CL{{\cal L}}

\def\CO{{\cal O}}

\def\CS{{\cal S}}
\def\CT{{\cal T}}

\newcommand{\half}{\frac{1}{2}}

\newcommand{\ten}[3]{{#1}^{#2}_{#3}}

\newcommand{\liexi}[1]{\mathcal{L}_{\xi}{#1}}
\newcommand{\lie}[1]{\mathcal{L}{#1}}

\def\ct{\cite}

\def\eq#1{(\ref{#1})}

\def\a{\alpha}
\def\b{\beta}

\def\G{\Gamma}
\def\d{\delta}

\def\ep{\epsilon}
\def\e{\eta}
\def\ph{\phi}
\def\Ph{\Phi}

\def\Ps{\Psi}

\def\L{\Lambda}
\def\m{\mu}

\def\r{\rho}

\def\S{\Sigma}
\def\ta{\tau}

\def\zt{\zeta}

\def\half{\frac{1}{2}}

\def\na{\nabla}

\def\pa{\partial}

\def\td#1{\tilde{#1}}

\title{\boldmath Extended Near Horizon Symmetries of Extremal BTZ Black Holes in 3D Massive Gravity
}

 \author[]{Debojyoti Ballav}
\author[]{and Shailesh Kulkarni}
\affiliation[]{Department of Physics, Savitribai Phule Pune University, \\Ganeshkhind, Pune, 411007, India}

\emailAdd{debojyoti.ballav@gmail.com}
\emailAdd{shailesh.kulkarni@unipune.ac.in}

\abstract{We study the asymptotic symmetries of near-horizon extremal BTZ black holes in higher derivative theories of gravity, such as New Massive Gravity and Topological Massive Gravity. By employing a particular boundary condition and the regularization prescription proposed earlier for the Einstein gravity, we demonstrate the existence of two centrally extended Virasoro algebras. The central charges evaluated within this framework are in agreement with their corresponding expressions evaluated at the spatial infinity. We also discuss the robustness of the regularization procedure by relating asymptotic and near-horizon geometries.}

\begin{document} 
\maketitle
\flushbottom
\section{Introduction}
\label{sec:intro}
A multitude of research endeavors have been performed to investigate asymptotic symmetries in $2+1$ dimensional gravity. The goal is to deepen our understanding of the duality between quantum field theories and black holes. Brown and Henneaux's pioneering research \cite{Brown:1986nw} has now become a crucial component in establishing a definitive link between the asymptotic symmetries of black holes in anti-de-Sitter (AdS$_3$) spacetime and the symmetry generators of two-dimensional conformal field theory (CFT). The study found that in an asymptotically AdS$_{3}$ spacetime, such as the Banados-Teitelboim-Zanelli (BTZ) black hole \cite{Banados:1992wn}, the algebra of conserved charges associated with diffeomorphisms satisfying specific asymptotic boundary conditions results in two copies of the centrally extended Virasoro algebras. This result was then utilized to calculate the microscopic entropy of the BTZ black hole \cite{Strominger:1997eq}. These findings have encouraged many researchers to explore dualities that extend beyond AdS/CFT.

A major development in this regard has been made in acquiring the dual description for a more realistic $3+1$ dimensional Kerr black hole, referred to as the Kerr/CFT correspondence \cite{Guica:2008mu}. The Kerr/CFT framework is based on the observation that the near-horizon extremal Kerr (NHEK) black hole has $SL(2,R)\times U(1)$ as the isometry group. On a given slice of fixed polar angle, NHEK describes three-dimensional AdS$_{2} \times S^{1}$ geometry, also known as warped AdS$_{3}$ (WAdS$_{3}$) \cite{Bardeen:1999px}. The asymptotic symmetry generators (ASG) that preserve the well-chosen boundary conditions result in the enhancement of the $U(1)$ sector and satisfy a Diff($S^1$) algebra. The associated asymptotically conserved charges lead to a centrally extended Virasoro algebra, thus having chiral CFT as a dual theory. The said correspondence has now been realized in various setups \cite{Lu:2008jk, Azeyanagi:2008kb,Isono:2008kx,Compere:2009dp, Hartman:2008pb,Wu:2009di,Rasmussen:2010sa,Peng:2009ty} including the black holes in higher derivative theories \cite{Azeyanagi:2009wf} (See \cite{Bredberg:2011hp, Compere:2012jk} for the extensive reviews on this subject.).    

Numerous efforts have been made to identify the appropriate boundary conditions that would permit the coexistence of two central charges ($c_L$ and $c_{R}$) near the horizon at extremality, with the expectation that this would yield insights into dual descriptions for black holes away from the extremality \cite{Azeyanagi:2008dk, Castro:2009jf,  Matsuo:2009sj, Matsuo:2010ut}. For instance, in \cite{Matsuo:2009sj}, a stronger set of boundary conditions was imposed that enhances the $SL(2, R)$ sector instead of $U(1)$ to the Virasoro level. In \cite{Matsuo:2010ut}, a new near-horizon limiting procedure for a five-dimensional Myers-Perry black hole was introduced to realize the dual two-dimensional CFT. There are proposals to understand extremal Kerr/CFT duality without actually going near the horizon using a stretched horizon formalism \cite{Carlip:2011ax}. However, this approach also tends to produce a chiral CFT.  

In \cite{Azeyanagi:2011zj}, a method is proposed that enhances both $U(1)$ as well as the $SL(2,R)$ symmetries. It is motivated by the fact that the geometry of the extremal black hole in the zero-entropy limit exhibits an emergent local AdS$_{3}$ structure \cite{Azeyanagi:2010pw}. However, the local AdS$_{3}$ has the same geometry as that of near-horizon extremal BTZ (NHEBTZ) black holes. It was shown in \cite{Balasubramanian:2009bg, deBoer:2010ac} that for NHEBTZ the diffeomorphisms satisfying the boundary conditions similar to \cite{Brown:1986nw} give rise to a chiral Virasoro algebra. Using this feature and introducing consistent boundary conditions, it was demonstrated that the ASG conforms to two sets of Diff($S^1$) algebras. Furthermore, following the specific regularization (deformation) prescription\footnote{In this paper, we shall use the words regularization and deformation interchangeably with the understanding that they describe the same physical situation.}, the algebra of asymptotic charges was shown to be isomorphic to two centrally extended Virasoro algebras, with non-zero $c_L$ and $c_R$ both remaining independent of the parameter dictating the regularization. Although the work of \cite{Azeyanagi:2011zj} was motivated by the Kerr/CFT correspondence, it is the structure of NHEBTZ that plays a crucial role in the proposed non-chiral extension of the symmetry algebra.

The majority of the previously mentioned works, which aimed to expand symmetry algebra, have concentrated their analyses on Einstein's gravity. It is generally anticipated that the Lagrangian will incorporate higher-derivative correction terms of the metric and other fields, assuming that Einstein's gravity serves as the low-energy effective theory of its ultraviolet completion, such as string theory. Compared to higher dimensions, gravity in $2+1$ dimensions offers significant insight into specific aspects of gravitational theory. It enables the exploration of diverse facets of black holes as it may encompass them and possesses a dual representation via the AdS/CFT correspondence. The inclusion of additional terms, such as higher derivative terms, enhances the physical content, as they facilitate the propagation of gravitons. One such model is Topologically Massive Gravity (TMG), proposed by Deser {\it et al.} \cite{Deser:1982vy}, which accommodates massive, parity-violating degrees of freedom. Alongside the BTZ solution, TMG presents various interesting black hole solutions \cite{Clement:1994sb, Bouchareb:2007yx}. Another well-studied example of higher derivative gravity in $2+1$ dimensions is New Massive Gravity (NMG), initially introduced as the parity-even counterpart of TMG \cite{Bergshoeff:2009hq}. NMG also allows for the propagation of massive gravitons and admits interesting black hole solutions \cite{Bergshoeff:2009aq, Clement:2009gq, Andringa:2009yc}. 

There are many instances where the asymptotic symmetry analysis has been conducted for the black holes in higher derivative theories of gravity, both at spatial infinity and at the near-horizon \cite{Compere:2007in, Hotta:2008yq,Compere:2008cv,Oliva:2009ip,Liu:2009kc,Anninos:2008fx,Compere:2009zj,Kim:2013qra} (for more recent works see \cite{Afshar:2015wjm, Ghodrati:2016vvf,Detournay:2016gao, Zwikel:2016smm, Adami:2017phg, Donnay:2020yxw, Liu:2021hvb,Liu:2022uox}). However, the non-chiral extension of near-horizon symmetries, as examined by Azeyanagi {\it et al.} \cite{Azeyanagi:2011zj}, remains unaddressed in the context of higher curvature theories. Two distinct aspects of this approach are crucial to its effectiveness. The initial consideration is the selection of the boundary condition that determines the structure of $Vir \times Vir$ of the asymptotic symmetry generators. Secondly, there are canonical asymptotic charges associated with these generators. The asymptotic charge algebra for left movers exhibits central extension, whereas the right movers give vanishing contributions. By applying a regularization procedure, it was shown that the Lie brackets among the right-moving asymptotic charges yield the correct central charge. Generally, in the computation of asymptotic conserved charges within Einstein gravity, one typically employs the ADT \cite{Deser:2002rt}, Wald-Iyer \cite{Iyer:1994ys}, or Barnich-Brandt \cite{Barnich:2001jy, Compere:2018aar} formalism. It is known that the expressions for conserved charges need to be improved for the gravity Lagrangian that uses the general higher-derivative terms \cite{Azeyanagi:2009wf}. In these instances, the extra contribution in the charge expression may disrupt the deformation process, resulting in a vanishing or divergent central extension term, indicating the necessity for further investigation. The purpose of this study is to present a non-chiral extension of the near-horizon asymptotic symmetries for the extremal BTZ black holes in the higher derivative theories of gravity like NMG and TMG. 

We organize the paper as follows. In section \ref{2.1}, we begin with the near-horizon extremal BTZ black hole in NMG. We first demonstrate that the boundary conditions and diffeomorphisms employed in conventional Kerr/CFT result in a left-moving, centrally extended Virasoro algebra. Section \ref{2.2} provides a detailed discussion of the non-chiral boundary condition \footnote{We will refer to the boundary condition of Kerr/CFT as the chiral boundary condition. The boundary condition presented in \cite{Azeyanagi:2011zj} is termed the non-chiral boundary condition.} and the regularization procedure.  The asymptotic symmetry charges associated with the left and right-moving ASG together with the regularization prescription lead to two copies of the centrally extended Virasoro algebras. In section \ref{3}, we consider the NHEBTZ in TMG and present the analysis for the regularized case only. Following a procedure similar to that in NMG, we show the coexistence of left and right central charges. In section \ref{4}, we comment on the possible relationship between the regularized asymptotic symmetry generators for NHEBTZ black holes and the corresponding one for asymptotically AdS$_{3}$ black holes. We summarize our results and discuss some future directions in the final section. In appendix \ref{extremallimit} we briefly discuss the near-horizon limiting procedure for the extremal BTZ black hole. Appendix \ref{asymptoticcharges} contains relevant expressions for the conserved charges in Barnich-Brandt-Compere (BBC) formalism.   

 
\section{Analysis for NMG}\label{2}

In this section, we consider the NHEBTZ black hole in NMG and study the asymptotic symmetries by implementing the chiral \cite{Guica:2008mu} as well as non-chiral \cite{Azeyanagi:2011zj} boundary conditions.

The action for the NMG is given by \cite{Bergshoeff:2009hq,Bergshoeff:2009aq}
\begin{align}\label{eq:NMGAction}
    \CA_{NMG} = \frac{1}{16\pi G_3} \int d^3x \sqrt{-g} \left( R - 2\lambda -  {\frac{1}{m^2}K} \right).
\end{align}
Where $ m $ represents the coupling constant and $ K $ is defined as: $ K \equiv R_{ab} R^{ab} - \frac{3}{8} R^2 $. The equations of motion for NMG theory are given by
\begin{align}\label{eq:nmgeqmot}
    R_{ab} -& \frac{1}{2} R g_{ab} + \lambda g_{ab} - \frac{1}{2 m^2} K_{ab} = 0, 
\end{align}
with
\begin{align*}
    K_{ab}=&-\frac{1}{2}\na^2R g_{ab}-\frac{1}{2}\na_{a}\na_b R+2\nabla^2R_{ab}-8{R_a}^c R_{cb}+\frac{9}{2}RR_{ab}+g_{ab}\bigg(3R_{cd}R^{cd}-\frac{13}{8}R^2\bigg).
\end{align*}
 The consequence of diffeomorphism invariance of the action imposes a constraint on $K_{ab}$, i.e., $\na_a K^{ab} = 0$, and taking the trace of \eq{eq:nmgeqmot}, we also have $m^2 R = K$. The stationary, circularly symmetric solutions are obtained by substituting the metric ansatz into the NMG theory. The reduced NMG Lagrangian then becomes $SL(2, R)$ invariant. By exploiting the local isomorphism between $SL(2,R)$ and $SO(2,1)$, the resultant equations of motion and Hamiltonian constraints can be expressed as vector algebraic equations that have either the BTZ black hole or WAdS$_{3}$ solution (see equation (3.17) in \cite{Clement:2009gq}). This $SL(2,R)$ reduction procedure has been used to generate various black hole solutions \cite{Andringa:2009yc}. Here we will focus only on the BTZ solution as given by metric \eq{eq:NonExtremeBTZ} with identification $\lambda = \Lambda - {\Lambda^2}/{4m^2} $ ($ \Lambda = -{1}/{\ell^2} $), $\ell$ being the AdS$_3$ radius. 

Our next task is to investigate near-horizon symmetries for the extremal BTZ black hole in NMG. 
For that, we take the near-horizon limit of the extremal BTZ metric.   
Details about this limiting procedure are given in appendix \ref{extremallimit}. By executing this procedure, we obtain the NHEBTZ metric 
\begin{align} \label{eq:extremalBTZnearhor} 
    ds^2=&\frac{\ell^2}{4}\bigg[-r^2 dt^2+\frac{dr^2}{r^2}+(d\ph-rdt)^2\bigg].
\end{align}
The $\ph$-direction is compactified as $(t,\ph)\sim (t,\ph+2\pi \eta)$, where $\eta$ is a constant fixed by the mass of the BTZ black hole as given in \eq{phiperodicity} $\e=2 r_h/\ell$ and $r_h=\ell\sqrt{M}$. 
The line element in \eq{eq:extremalBTZnearhor} has the structure of AdS$_2 \times S^1$ with $ SL(2, R) \times U(1) $ as the isometry group  generated by the  Killing vectors
\begin{align}\label{isometries-NHEBTZ}
    \xi_1=&\pa_t, \ \xi_2=\pa_\ph, \ \xi_3= t\pa_t-r\pa_r, \ \xi_4= -\bigg(\frac{t^2}{2}+\frac{1}{2 r^2}\bigg)\pa_t + rt\pa_r-\frac{1}{r}\pa_\ph.
\end{align}
We now study the near-horizon symmetries for the extremal geometry \eq{eq:extremalBTZnearhor} by imposing appropriate boundary conditions on the metric perturbation or excitations.

\subsection{Chiral boundary condition}\label{2.1}

One possible choice is the 2+1 dimensional analogue of the Kerr/CFT boundary condition in \ct{Guica:2008mu} which reads as $g_{ab} = \bar{g}_{ab} + h_{ab}$ with 

\begin{align}\label{stromingerBC} 
h_{ab} \sim \begin{pmatrix}
\CO({r^2}) & \CO({r^{-2}}) &  \CO(1)\\
& \CO({r^{-3}}) & \CO({r^{-1}}) \\
   &  & \CO(1) \\
\end{pmatrix},
\end{align}
where the order of the coordinates in the matrix is taken to be $(t,r,\ph)$. The most general diffeomorphism that preserves this boundary condition will have a form
\begin{align}
    \xi= f(\ph)\pa_\ph-rf'(\ph)\pa_r,
\end{align}
since $\ph\sim \ph+2\pi\eta$ it is convenient to choose a Fourier basis for the function $f(\ph)$ given by $f_n(\ph)=-\eta e^{-in\ph/\eta}$ and the generators decomposed in that basis are
\begin{align}\label{NHEBTZgenrfourier}
\xi_{n} = -\eta\; e^{-in\ph/\eta} \pa_\ph -  i n r\;e^{-in\ph/\eta} \pa_r.     
\end{align}
Lie bracket between these asymptotic generators satisfies the Diff($S^1$) or Witt algebra :  
\begin{align}
    \{\xi_n,\xi_m\}= \lie{_{\xi_n}(\xi_m}) =i(m-n)\;\xi_{m+n}.
\end{align}
As mentioned in \cite{Guica:2008mu}, the enhancement in the symmetry algebra is coming from the $U(1)$ sector of the isometry group $SL(2,R) \times U(1)$.

We now compute the central extension term corresponding to the generator
\eq{NHEBTZgenrfourier} by using the BBC formalism
\cite{Barnich:2001jy}, and the relevant formulas are given in
appendix \ref{asymptoticcharges}. For the central extension, 
we use \eq{surfacechargeshen} with $h_{ab}=\CL_{\xi}\bar{g}_{ab}$ 
where the diffeomorphism $\xi$ is given in \eq{NHEBTZgenrfourier} and thus we arrived at

\begin{align}\label{eq:ktrforleftdiffnonchiral}
  k^{tr}_{\xi_q}[\CL_{\xi_p}\bar{g};\bar{g}]&=\frac{1}{8\pi G_3} \bigg(\frac{i e^{-i(p+q)\ph/\eta}}{40 \ell m^2 \eta}\bigg)\; \;\bigg\{(3p^3-7p^2q+10 p)+10\ell^2 m^2(p^2q-p^3-2p)\bigg\}.
\end{align}
 Integrating over $\phi \in [0,2\pi\eta]$ on the constant $t-r$ surface $\S$, leads to the 
 central extension term
 \begin{align}\label{eq:centraltermchiral}
      C_{\xi_{q},\xi_{p}} = \frac{i\ell}{(4G_3)(40\ell^2 m^2)}\;\bigg\{(3p^3-7p^2q+10 p)+10\ell^2 m^2(p^2q-p^3-2p)\bigg\}\delta_{p,-q}.
 \end{align}

 \noindent Shifting the generators as $L_p=Q_{\xi_p}+\delta_{p,0}\times(const)$ 
 and elevating it to the quantum level (with the corresponding replacement 
 $\{.\; , . \; \} \rightarrow -i[.\; , . \;]$) we get a copy of the centrally 
 extended Virasoro algebra:
 
\begin{align}\label{virasoroNMGstrominer}
    [L_p,L_q]=&(p-q)L_{p+q}+\frac{c}{12}p(p^2-1)\delta_{p,-q},
    \end{align}
 where the central charge $c$ is derived from the coefficient of $p^3$ term in \eq{eq:centraltermchiral} with the identification 
  \begin{align}\label{eq:3.21}
        c= 12 i C_{\xi_{-p},\xi_{p}}|_{p^3} =\frac{3\ell}{2G_3}\left(1-\frac{1}{2m^2\ell^2}\right).
    \end{align}
 A similar result for NMG was obtained in the literature \cite{Kim:2013qra} using the stretched-horizon approach of Carlip \cite{Carlip:2011ax}. Therefore, we see that the chiral (Kerr/CFT-type) boundary condition (\ref{stromingerBC}) results in an enhancement in the symmetry algebra among the asymptotic symmetry generators, and the corresponding field representation (quantum charges) satisfies a centrally extended Virasoro algebra. 

  \subsection{Non-chiral boundary condition}\label{2.2}
  
 We now consider the boundary condition of \cite{Azeyanagi:2011zj} that will lead to the non-chiral extension of the asymptotic symmetries of NHEBTZ  
 \begin{align}\label{nonchiralBC}
g_{ab} = \bar{g}_{ab} + h_{ab}, \quad h_{ab} \sim \begin{pmatrix}
\CO(1) & \CO(r^{-1}) & \CO(1)\\
&\CO({r^{-3}})&\CO(r^{-1}) \\
   &  & \CO(1) \\
\end{pmatrix}.
\end{align}
The above boundary condition is very specific for the metrics where $\bar{g}_{tt} = 0$ or at best a constant. The ASG that preserves the boundary condition \eq{nonchiralBC} given by
\begin{align}\label{nonchiraldiffeo}
    \zt=\left[\ta(t)+ \CO(r^{-1})\right]\pa_t+\left[-r(\ta'(t)+\ep'(\ph))+ \CO(1)\right]\pa_r + \left[\ep(\ph)+ \CO(1/r) \right] \pa_\ph,
\end{align}
where $\ta(t)$ and $\ep(\ph)$ are arbitrary functions of $t$ and $\ph$, respectively and prime on $\ta(t)$ and $\ep(\phi)$ denotes derivative with respect to the corresponding  arguments. By evaluating $\lie{_{\zt}\bar{g}_{ab}}$ we find
\begin{align}\label{Liedernonchiraldiffeo}
\lie{_{\zt}\bar{g}_{tt}}=\lie{_{\zt}\bar{g}_{t\phi}}=\lie{_{\zt}\bar{g}_{\phi\phi}} = \CO(1)  \ ; \ \lie{_{\zt}\bar{g}_{tr}} = \lie{_{\zt}\bar{g}_{r\phi}}=\CO(1/r)  \ ; \ 
 \lie{_{\zt}\bar{g}_{rr}} = \CO(1/r^3).
 \end{align}
Thus, the diffeomorphism in \eq{nonchiraldiffeo} is an asymptotic Killing vector for the geometry \eq{eq:extremalBTZnearhor}. Notice the appearance of two arbitrary functions $\ta(t)$ and $\ep(\phi)$ in $\zt$. As before, we take $\ep(\ph)$ to be periodic. But unlike in the case for the boundary condition \eq{stromingerBC}, the diffeomorphism that preserves \eq{nonchiralBC} also contains an arbitrary time function, $\ta(t)$. As we shall see below, this freedom enables us to construct left as well as right-moving asymptotic Killing vectors. After setting the periodicity for $t$ as $t \sim t + \a $, where $\a $ is any positive constant, and picking the Fourier bases for $\ta(t)$ and $\ep(\ph)$ as
\begin{align}\label{modefunctn}
\ta_n(t) = -\frac{\a}{2\pi} e^{-2\pi in t/\a}, \quad \ep_n(\ph) = -\eta\; e^{-in\ph/\eta},
\end{align}
we expand the ASG generators in the above bases  

\begin{subequations}\label{nonchiraldiffeofourier}
\begin{align}
    \zt^+_{n}=&-\eta\; e^{-in\ph/\eta}\; \pa_\ph -inr e^{-in\ph/\eta}\;\pa_r, \label{eq:3.8b}\\
     \zt^{-}_{n}=&-\frac{\a}{2\pi}e^{-2\pi int/\a} \pa_t-inr e^{-2\pi int /\a} \pa_r\label{eq:3.8a},
\end{align}
\end{subequations}
where $(+)$ corresponds to left-moving and $(-)$ for right-moving sectors, respectively. The Lie brackets among these diffeomorphisms yield 
\begin{align}\label{nonchiralvirasoro}
\{\zt_m^{\pm}, \zt_n^{\pm} \}  =& -i(m - n) \zt_{m+n}^{\pm}, \quad
  \{\zt_m^-, \zt_n^+ \} =  \ 0. 
\end{align}

By comparing the diffeomorphisms \eq{nonchiraldiffeofourier} and \eq{NHEBTZgenrfourier}, it is evident that the original $U(1)$ sector is enhanced to the Virasoro level. Furthermore, there is an enhancement of the $SL(2,R)$ sector that can be interpreted as a non-chiral extension of the Kerr/CFT duality. 

We now compute the central extension term for $\zt^+$ and $\zt^-$ generators. For the left-moving part, the expression for $k^{tr}_{\zt^+_q}[\CL_{\zt^+_p}\bar{g};\bar{g}] $ will have the same form as in \eq{eq:ktrforleftdiffnonchiral}. Thus, the central extension term $C_{\zt^{+}_{p},\zt^{+}_{q}}$ will be the same as that given in \eq{eq:centraltermchiral}. We then identify the left-moving central charge as  
 \begin{align}\label{eq:leftcentralchgargenonchiral}
        c_{L} = \frac{3\ell}{2G_3}\left(1-\frac{1}{2m^2\ell^2}\right).
    \end{align}
 Performing a similar analysis for the right-moving generator $\zt^{-}$ we get 
 the expression for $(t,r)$ component of BBC potential 
\begin{align}\label{eq:ktrforrightdiffnonchiral}
   k^{tr}_{\zt^-_q}[\CL_{\zt^-_p}\bar{g};\bar{g}]=\CO \left(\frac{1}{r^3}\right).
\end{align}
Notice that the leading term in the above expression is proportional to $1/r^3$. Consequently, the corresponding central extension term, $C_{\zt^{-}_{p},\zt^{-}_{q}}$, which is $\phi$ integration over the $\Sigma$ at infinity vanishes, yielding the trivial right-moving central charge $c_R$. To sum up, we have found the left ($\zt^{+}_{n}$) and right ($\zt^{-}_{n}$) asymptotic symmetry generators that preserve \eq{nonchiralBC} and the conserved charges $Q_{n}$ corresponding to both left and right moving modes satisfy the Virasoro algebra, but only the left moving sector gets centrally extended.

The above algebraic asymmetry between left- and right-moving modes can be understood by looking at the near-horizon structure of the extremal BTZ metric. The choice of $t=const$ hypersurface $\Sigma$ to calculate the charges in this naive prescription becomes problematic because here $\Sigma$ is not a spacelike surface. It becomes null at the conformal boundary $\CB$, which can be seen explicitly by considering the metric \eq{eq:extremalBTZnearhor} on $\CB$ by taking $dr=0$ and $r\to \infty$ limit. In that case, the metric will be $ds^2|_{\CB}=-dtd\ph$. Clearly, this metric vanishes for $dt = 0$, so the equal-time surface $\Sigma$ is null at the boundary. 

To overcome this problem we first consider a general linear transformation $(t,\phi)\rightarrow (t',\ph')$ as 
\begin{align}\label{newtphicoorgeneral}
    t' = \beta_{1}t + \beta_{2} \ph \ ; \ \ph' = \beta_{3}t + \beta_{4}\ph,
\end{align}
where $\beta_{1}, \cdots, \beta_{4}$ are real constants. In $(t',\ph')$ system the metric $ds^{2}_{reg}$,  takes the form 
\begin{align}\label{genregmetric}
\begin{split}
ds^2_{reg} =& \frac{\ell^2}{4}\frac{dr^2}{r^2}+ \frac{\ell^2}{4}\frac{1}{(\beta_{1}\beta_4 - \b_2\b_3)^{2}}\bigg[ (\beta^{2}_{1} + 2 r \beta_1 \beta_2)d\ph'^{2} + (\beta^{2}_{3} + 2 r \beta_3 \beta_4)dt'^{2}\\ 
-&  2dt'd\ph' \bigg\{r(\b_1\b_4+\b_2\b_3) - \b_1\b_3\bigg\} \bigg].
\end{split}
\end{align}
Thus, the metric on the conformal boundary $(r\rightarrow \infty)$ is now given by
\begin{align}\label{genregmetric2}
    ds^{2}_{reg}|_{\CB} =  \frac{2 r \beta_1 \beta_2}{(\beta_{1}\beta_4 - \b_2\b_3)^{2}}d\ph'^{2}.
\end{align}
For it to be finite and positive definite on $\CB$ implies $\b_{1}\b_4 \ne \b_2\b_3 \ \ \text{and} \ \ \b_1 \ne 0 \ ;   \b_2 \ne 0$.
The choice that preserves the original compactification, that is, $ (t',\ph')\sim (t',\ph'+2\pi \eta)$, fixes $\b_4=1$. Demanding $g_{t't'}$ to be zero for the requirement of the boundary condition \eq{nonchiralBC} to work, we fix $\b_3=0$. The other two parameters can be adjusted such that the effective ratio $\b_2/\b_1$ can be taken as a regularization (deformation) parameter. As a special case, if we take $\b_1=1$ and rename $\b_2=\b$, we get the situation as in \cite{Azeyanagi:2011zj}. With this choice the transformation becomes 
\begin{align}\label{newtphicoor}
    t'=t+\b\ph, \:\: \ph'=\ph.
\end{align}
 In the computations involving asymptotic charges, we initially approach the limit $r\to \infty$ and then take $\b\to 0$. The geometry \eq{eq:extremalBTZnearhor} in the new $(t',r,\ph')$ coordinates  transforms into 
\begin{align}\label{eq:extremalBTZnearhorreg}
     ds^2_{reg}=\frac{\ell^2}{4}\left[\frac{dr^2}{r^2}-2rdt'd\ph'+(1+2\b r)d\ph'^2\right].
\end{align}
Above metric still admits $SL(2,R)\times U(1)$ isometry group generated by 
\begin{align}\label{isometries-NHEBTZreg}
\begin{split}
    \zt_1=&\pa_{t'} \ , \ \zt_2=\b \pa_{t'}+\pa_{\ph'} \ , \ \zt_3= t\pa_{t'}-r\pa_r,\\
    \zt_4=& -\bigg(\frac{1}{2}(t'-\b\ph')^2+\frac{1}{2 r^2}+\frac{\b}{r}\bigg)\pa_{t'} + r(t'-\b\ph')\pa_r-\frac{1}{r}\pa_{\ph'},
\end{split}
\end{align}
which can be easily confirmed by transforming \eq{isometries-NHEBTZ} according to \eq{newtphicoor}. Taking the metric \eq{eq:extremalBTZnearhorreg} as $\bar{g}_{ab}$ and imposing the boundary condition \eq{nonchiralBC}, one can easily obtain the asymptotic symmetry generator $\zt$ as in \eq{nonchiraldiffeo}. To obtain the Fourier decomposition of the generators, we first note that the $t'$ has the periodicity $t' \sim t +  \beta(\ph+2\pi\eta) = t' + 2\pi\beta\eta$. Then the Fourier mode generators are given by
\begin{subequations}\label{nonchiraldiffeofourierreg}
\begin{align}
    \zt^+_n &= - e^{-in\ph'/\eta} \big[\b \eta\; \pa_{t'}+inr\pa_r+\eta\; \pa_{\ph'}\big],
    \label{nonchiraldiffeofourierregleft}\\
     \zt^-_n &= - e^{-\frac{in}{\beta\eta}(t'-\beta\phi')}\big[\b \eta\; \pa_{t'}+inr\pa_r\big]. \label{nonchiraldiffeofourierregright}
\end{align}
\end{subequations}
We refer to the diffeomorphisms in \eq{nonchiraldiffeofourierreg} as the regularized or deformed asymptotic symmetry generators. Computing the Lie bracket among these generators we see that they obey the same algebra as in \eq{nonchiralvirasoro}. Next, we compute the algebra among the asymptotic conserved charges corresponding to the deformed generators \eq{nonchiraldiffeofourierreg}. Using \eq{surfacechargeshen}, the $(t'r)$ component of the BBC potential for the left ($\zt^+_n$) and right ($\zt^-_n$)  generators are given by 

\begin{align}\label{eq:ktrforleftdiffnonchiralreg}
  \td{k}^{t'r}_{\zt^+_q}[\CL_{\zt^+_p}\bar{g}';\bar{g}']=& \frac{1}{8\pi G_3}\bigg(\frac{i}{40\ell m^2 \eta}\bigg)\;e^{-i(p+q)\ph'/\eta}\bigg\{(3p^3-7p^2q+10 p)\nonumber\\
  +&10\ell^2 m^2(p^2q-p^3-2p)\bigg\}  
  + \CO(r^{-1}),
\end{align}
\begin{align}\label{eq:ktrforrightdiffnonchiralreg}
    \td{k}^{t'r}_{\zt^-_q}[\CL_{\zt^-_p}\bar{g}';\bar{g}']=& \frac{1}{8\pi G_3}\bigg(\frac{i}{40\ell m^2 \eta}\bigg)\;e^{\frac{-i(p+q)}{\beta\eta}[t'-\beta\ph']}\bigg\{(3p^3-7p^2q)+10 \ell^2 m^2(p^2 q-p^3)\bigg\} \nonumber \\ + &  
   \  \CO(r^{-1}).
\end{align}
It is important to note that $ \td{k}^{t'r}_{\zt^+_q}[\CL_{\zt^+_p}\bar{g}';\bar{g}']$ is equivalent to $k^{tr}_{\zt^+_q}[\CL_{\zt^+_p}\bar{g};\bar{g}]$ (i.e., prior to regularization). On the other hand, we get a non-zero $(t'r)$ component for the BBC potential that corresponds to the right-moving generators. The fact that the central extension terms for left-moving generators remain unaffected and that the right-moving generators give nonvanishing results post-regularization can be explained by looking at the coordinate transformations $(t,\ph) \rightarrow (t',\ph')$. Under this, $k^{tr}_{\zt}\rightarrow \td{k}^{t'r}_{\zt} = k^{tr}_{\zt} + \beta k^{\phi r}_{\zt}$. Here $\zt$ appearing on the left side stands for regularized generators \eq{nonchiraldiffeofourierreg}, while the one on the right-hand side is given by \eq{nonchiraldiffeofourier}.  The $(\ph,r)$ components of unregularized BBC potential for the left and right moving modes can be worked out as $k^{\ph r}_{\zt^+_q}[\CL_{\zt^+_p}\bar{g};\bar{g}] = 0,$ and
\begin{align}\label{eq:kphirforleftrightdiffnonchiral}
\begin{split}
k^{\ph r}_{\zt^-_q}[\CL_{\zt^-_p}\bar{g};\bar{g}]=
  \frac{1}{8\pi G_3} \bigg(\frac{i\pi e^{-2\pi i(p+q)t/\alpha} }{20 \ell \alpha m^2}\bigg)\; \;\bigg\{(3p^3-7p^2q)+10 \ell^2 m^2(p^2 q-p^3)\bigg\}. 
\end{split}
\end{align}
Now, replacing $\alpha$ by $2\pi\eta \beta$ and $t= t'- \beta\ph'$ in the second of the above equation, and noting \eq{eq:ktrforrightdiffnonchiral}, we see that the combination $k^{tr}_{\zt^-_q}[\CL_{\zt^-_p}\bar{g};\bar{g}] + \beta k^{\ph r}_{\zt^-_q}[\CL_{\zt^-_p}\bar{g};\bar{g}]$ precisely gives the right-hand side of \eq{eq:ktrforrightdiffnonchiralreg}. On the other hand, since the $(\ph, r)$ component of the unregularized BBC potential for left-moving modes vanishes identically, the entire contribution comes from $k^{tr}_{\zt^+}$ (see \eq{eq:ktrforleftdiffnonchiral}) producing the right-hand side of \eq{eq:ktrforleftdiffnonchiralreg}. Integrating \eq{eq:ktrforleftdiffnonchiralreg} and \eq{eq:ktrforrightdiffnonchiralreg} on the constant $t'-r$ surface,  we get non-vanishing central extension terms for the left and right moving generators. After carrying out the usual quantization procedure, we see that the quantum  generators satisfy two centrally extended Virasoro algebras 
\begin{align}\label{nonchiralvirasoroNMG}
    [L^{+}_p,L^{+}_q]=(p-q)L^{+}_{p+q}+\frac{c_{L}}{12}\;p(p^2-1)\delta_{p,-q} \;,\\
    [L^{-}_p,L^{-}_q]=(p-q)L^{-}_{p+q}+\frac{c_{R}}{12}p(p^2-1)\delta_{p,-q}\;,
    \end{align}
with the central charges 
\begin{align}\label{eq:leftrightcentralchgargenonchiralreg}
        c_{L} = c_{R}=\frac{3\ell}{2G_3}\left(1-\frac{1}{2m^2\ell^2}\right).
    \end{align}
 Consequently, our analysis indicates the presence of nonzero left and right central charges for the NHEBTZ black hole in NMG. It is noteworthy that both central charges remain unaffected by the periodicity of $\ph$. This feature consistently holds for the Kerr/CFT correspondence examined under various boundary conditions, \cite{Compere:2012jk}.
Furthermore, the central charges are independent of the parameter $\b$. 
Our results for $c_{L}$ and $c_{R}$ align precisely with those derived from the well-known Brown-Henneaux boundary condition at $r\rightarrow \infty$ for BTZ black holes in NMG \cite{MahdavianYekta:2020lde, Chen:2013aza}.

  We now mention the thermodynamical aspects of extremal geometry in NMG. Using the general formula for the BBC potential \eq{surfacechargeshen} for time-like and rotational Killing vectors for the metric given in \eq{NHEBTZ} one gets the mass and angular momentum \cite{Nam:2010ub}
\begin{align}\label{eq:MassAngNMG}
\begin{split}
    M_{NMG} =&\frac{r_h^2}{4 G_3 \ell^2}\left( 1-\frac{1}{2m^2\ell^2} \right)=\frac{\e^2}{16G_3} \left( 1-\frac{1}{2m^2\ell^2} \right) , \\
    J_{NMG}=&\frac{r_h^2}{4 G_3 \ell}\left( 1-\frac{1}{2m^2\ell^2} \right)= \frac{\e^2\ell}{16G_3}\left( 1-\frac{1}{2m^2\ell^2} \right),
\end{split}
\end{align}
where in each of the above equations we have expressed the horizon radius in terms of periodicity parameter $\eta$ by the relation $r_{h} ={\e\ell}/{2}$ (see below \eq{phiperodicity}). The microscopic entropy of the extremal BTZ black hole near the horizon is obtained using Cardy's formula 
\begin{align}\label{eq:cardyformula}
\begin{split}
    \CS_{CFT}=2\pi\sqrt{\frac{c_L (M\ell +J)}{12}}+2\pi\sqrt{\frac{c_R (M\ell -J)}{12}}
    =\frac{\pi\e \ell}{4G_3}\left(1-\frac{1}{2m^2\ell^2}\right).
\end{split}
\end{align}
Where we have used the extremality condition $M\ell =J$. This calculation of the near-horizon entropy is exactly in agreement with the  Wald entropy for the BTZ black hole in NMG \cite{MahdavianYekta:2020lde}.

 To determine the Frolov-Thorne temperatures for the left $(T_{L})$ and right $(T_{R})$ sectors, respectively, we employ the methods in \cite{Chen:2012mh}. In the canonical ensemble, Cardy's formula in terms of the dimensionless CFT temperatures for left $(\CT_L)$  and right $(\CT_R)$ modes can be written as
\begin{align}\label{CardyformulaCan}
    \CS_{CFT}=\frac{\pi^2}{3} c_L \CT_L+ \frac{\pi^2}{3}c_R \CT_R.
\end{align}
\noindent A comparison of \eq{CardyformulaCan} with \eq{eq:cardyformula} reveals that $\CT_L$ and $\CT_R$ can be expressed in terms of mass, angular momentum, and central charge as
\begin{align}\label{DimlessFTtemp}
    \CT_{L,R}=\frac{1}{\pi}\sqrt{\frac{3(M\ell\pm J)}{c_{L,R}}}.
\end{align}
 Therefore, from \eq{DimlessFTtemp} using \eq{eq:MassAngNMG} and \eq{eq:leftrightcentralchgargenonchiralreg} 
we get $ \CT_L={r_h}/{\pi \ell}={\e}/{2 \pi};\:\CT_R=0$.
To get the dimensionful temperatures $T_{R,L}$  we note that the only length scale that is present in the BTZ geometry is $\ell$ and thus they related as $\CT_{L,R}=\ell\;T_{L,R}$. So we arrive at 
\begin{align}\label{eq:LRTemperature}
     T_L=\frac{r_h}{\pi\ell^2}=\frac{\e}{2 \pi \ell}, \quad T_R=0.
\end{align}


\section{Analysis for TMG }\label{3}

This section is devoted to studying the asymptotic symmetry analysis of the NHEBTZ black hole in the topologically massive gravity (TMG). We show the existence of two nonvanishing central charges $c_{L}$ and $c_{R}$. The TMG is represented by \cite{Deser:1982vy}
\begin{align}\label{eq:3.35}
    \CA_{TMG}=\frac{1}{16\pi G_3}\int d^3x \sqrt{-g} (R-2\L) + \frac{1}{16\pi G_3 \m} \CA_{CS},
\end{align}
 where $\CA_{CS}$ is the Chern-Simons term
 \begin{align}\label{eq:3.36}
     \CA_{CS}=\half\int d^3x\sqrt{-g}\; \ep^{abc} \;\G^d_{ce} \left(\pa_a\G^e_{db}+\frac{2}{3}\G^e_{am}\G^m_{bd} \right),
 \end{align}
with $\L=-1/\ell^2$. The sign chosen in front of the action is to make the energy of BTZ black hole positive for large $\m$. 
The equations of motion for this action are given by,
\begin{align}\label{eq:3.37}
    G_{ab}+\frac{1}{\m}\;\CC_{ab}=0,
\end{align}
where $G_{ab}$ is the Einstein tensor $G_{ab}\equiv R_{ab}-\half g_{ab} R+\L g_{ab}$ and $\CC_{ab}$ is the Cotton tensor $\CC_{ab}\equiv {\ep_{a}}^{cd}\na_c\left(R_{db}-\frac{1}{4}g_{db}R \right)$.
The Cotton tensor involved the third derivative of the metric. The convention chosen for the Levi-civita tensor $\ep^{abc}$ is $\ep^{tr\ph}=\frac{-1}{\sqrt{-g}}$ in the coordinates $(t,r,\ph)$.

    The asymptotic symmetries for the BTZ black hole \eq{eq:NonExtremeBTZ} at spatial infinity have been analyzed in great detail in the literature. It was shown that the Brown-Henneaux boundary conditions lead to two unequal central charges in the left and right sectors \cite{Hotta:2008yq}. 
   In addition to the BTZ solution, TMG also admits other interesting possibilities such as spacelike vacua and null warped vacua \cite{Anninos:2008fx}.  
   The asymptotic symmetry analysis for such warped geometries was carried out in \cite{Compere:2008cv, Compere:2009zj}. 
   
   We are interested in asymptotic symmetries for the NHEBTZ black hole metric \eq{eq:extremalBTZnearhor}. Since the procedure will be the same as in section \ref{2}, we confine ourselves to the non-chiral boundary condition and present only the relevant expressions for the central extension terms for the metric \eq{eq:extremalBTZnearhor} and its regularized counterpart \eq{eq:extremalBTZnearhorreg} \footnote{As in the case of NMG, we can show that the asymptotic charges corresponding to the boundary condition \eq{stromingerBC} for TMG produce one (left-moving) central charge.}. Similar to the section \ref{2.2}, we compute asymptotic charges and central extension terms using the algorithm given in appendix \ref{asymptoticcharges}. For TMG we need to take the first three terms of the BBC potential expression \eq{surfacechargeshen}. Using $h_{ab}=\CL_{\xi}\bar{g}_{ab}$ and substituting the expressions for the left $(\zt^{+}_{n})$ and right $(\zt^{-}_{n})$ symmetry generators in \eq{nonchiraldiffeofourier} we arrive at
\begin{align}\label{eq:TMGktrforleftrightdiffnonchiral}
\begin{split}
    k^{tr}_{\zt^+_{q}}[\CL_{\zt^+_p}\bar{g};\bar{g}]&=-\frac{1}{8\pi G_3} \bigg(\frac{i}{4\m \eta}\bigg)\; e^{-i(p+q)\ph/\eta}\;[(2p-p^2q+p^3)(1+\ell\m)]+\CO(r^{-1}), \\
    k^{tr}_{\zt^-_{q}}[\CL_{\zt^-_p}\bar{g};\bar{g}]&=\CO \left(\frac{1}{r^2}\right).
\end{split}
\end{align}
As before, integrating over $\ph \in [0,2\pi\eta]$ yields the central extension 
\begin{align}
C_{\zt^{+}_{q},\zt^{+}_{p}} = -\frac{i(1+\ell \mu)}{16\mu G_3}[2p-q p^2 + p^3] \delta_{p,-q}\;, \quad
C_{\zt^{-}_{q},\zt^{-}_{p}} = 0.
\end{align}

\noindent Thus, only the left-moving Virasoro algebra is centrally extended  with the central charge 
\begin{align}\label{eq:TMGcentralChargeLeft}
c_{L} = 12 i C_{\zt^{+}_{-p},\zt^{+}_{p}}|_{p^3} = \frac{3\ell}{2G_3}\left(1+\frac{1}{\mu\ell}\right).
\end{align}

As explained in section \ref{2.1}, the fact that the metric on the conformal boundary becomes null is reflected in the vanishing of the central extension term for the right-moving generators. Implementing the same regularization procedure through the transformations \eq{newtphicoor} and using the generators \eq{nonchiraldiffeofourierreg}, we obtain the regularized BBC potential for left-moving and right-moving generators as
\begin{subequations}\label{eq:TMGktrforrightdiffnonchiralreg}
\begin{align}
\td{k}^{t'r}_{\zt^+_{q}}[\CL_{\zt^+_p}\bar{g}';\bar{g}']&=-\frac{1}{8\pi G_3} \bigg(\frac{i }{4\m \eta}\bigg)\; e^{-i(p+q)\ph'/\eta}\;[(2p-p^2q+p^3)(1+\ell\m)]+\CO(r^{-1}), \label{eq:TMGktrforrightdiffnonchiralregL} \\
    \td{k}^{t'r}_{\zt^-_{q}}[\CL_{\zt^-_p}\bar{g}';\bar{g}']&=-\frac{1}{8\pi G_3}\bigg(\frac{i}{4\m \eta}\bigg)\;e^{\frac{-i(p+q)}{\beta\eta}[t'-\beta\ph']}[(p^2q-p^3)(1-\ell\m )]+\CO(r^{-1}), \label{eq:TMGktrforrightdiffnonchiralregR}
\end{align}
\end{subequations}
with the central charges
\begin{align}\label{centralchargeregTMG}
   c_{L} = \frac{3\ell}{2G_3}\left(1+\frac{1}{\mu\ell}\right) \ ;  c_R=\frac{3\ell}{2G_3}\left(1-\frac{1}{\mu\ell}\right). 
\end{align}
Thus, by exclusively utilizing near-horizon structures, we have successfully obtained the Virasoro algebras for left and right movers with distinct central charges. Our above expressions for $c_L$ and $c_R$ match exactly with the one calculated using Brown-Henneaux conditions at the spatial infinity of BTZ black hole \cite{Hotta:2008yq}. 
Note that since $\td{k}^{t'r}_{\zt} = k^{tr}_{\zt} + \beta k^{\phi r}_{\zt}$ the term which gives the non-zero contribution for the right-movers comes actually from the $(\phi,r)$ component of unregularised BBC potential 
\begin{align}\label{eq:TMGkphirforleftrightdiffnonchiral}
  k^{\ph r}_{\zt^-_q}[\CL_{\zt^-_p}\bar{g};\bar{g}]&=-\frac{1}{8\pi G_3}\bigg(\frac{i\pi}{2\m \alpha}\bigg)\;e^{-2\pi i(p+q)t/\a}[(p^2q-p^3)(1-\ell\m)]+\CO(r^{-1}).
\end{align}
Carrying out the replacements for $\alpha$ and $t$ as before, we recover \eq{eq:TMGktrforrightdiffnonchiralregR}. 
In the case of TMG the mass and angular momentum can be calculated using either BBC  or the method prescribed in \cite{Kim:2013cor} and the results are, 
\begin{align}\label{eq:MassAngTMG}
 \begin{split}
     M_{TMG}=\frac{\e^2}{16G_3}\left(1+\frac{1}{\m\ell}\right), \quad J_{TMG}=\frac{\e^2\ell}{16G_3}\left(1+\frac{1}{\m\ell}\right).  
 \end{split}
\end{align}
Using \eq{eq:cardyformula} we can determine the entropy as 
\begin{align}\label{TMGentropy}
    S_{CFT}=&\frac{\pi r_h}{2 G_3}\bigg(1 +\frac{1}{\m\ell} \bigg)=\frac{\pi\e \ell}{4G_3}\bigg(1 +\frac{1}{\m\ell} \bigg).
\end{align}
The Frolov-Thorne temperature only depends on $\e$ and $\ell$ and it will remain the same as calculated before in \eq{eq:LRTemperature}. 


\section{Relating asymptotic AdS$_3$ and deformed NHEBTZ}\label{4}
From our discussions of section \ref{2.2} and section \ref{3}, it is clear that the diffeomorphisms respecting the non-chiral boundary conditions form the $Vir\times Vir$ algebra, and the corresponding asymptotic charges, when properly deformed (regularized), lead to non-zero $c_L$ and $c_R$. Such a deformation is necessary to circumvent the null-like nature of the boundary (where the conserved charges are evaluated). We now argue that the procedure, when applied to cases where the boundary is already spacelike, does not change the pre-existing symmetry algebra. For concreteness, we consider the asymptotically AdS$_{3}$ black holes given by the metric
\begin{align}\label{eq:AdSMetric}
 ds^2= -\frac{R^2}{\ell^2}  dT^2 +\frac{\ell^2}{R^2} dR^2 + R^2 d\Ph^2. 
\end{align}
We impose the Brown-Henneaux boundary condition \cite{Brown:1986nw}
\begin{align}\label{eq:BraHannBoundaryCond}
g_{ab} = \bar{g}_{ab} + h_{ab} \sim\begin{pmatrix}
-\frac{R^2}{\ell^2}+ \CO(1) & \CO(r^{-3}) & \CO(1)\\
& \frac{\ell^2}{R^2} + \CO({r^{-4}})&\CO(r^{-3}) \\
   &  & R^{2}+ \CO(1) \\
\end{pmatrix}.
\end{align}
The diffeomorphisms that preserve it are given by
\begin{align}\label{eq:RegBrownGenerators}
  \varsigma^{\pm}_{p} = e^{-ip(T/\ell \pm \Ph)}
  \bigg[\frac{\ell}{2}\big(1-\frac{\ell^2 p^2}{2R^2}\big)\partial_{T} \pm 
  \frac{1}{2}\big(1+\frac{\ell^2 p^2}{2R^2}\big)\partial_{\Ph} - \frac{ipR}{2}\partial_{R}
  \bigg].
\end{align}
As is well known, the asymptotic analysis for this boundary condition when carried out at spatial infinity, yields $c_{L,R} = 3\ell/2G_3$ for the Einstein gravity. 
Suppose that we employ the regularization prescription by:  $ T'=T+\Upsilon \Ph, \quad \Ph'=\Ph$ where $\Upsilon$ has the connotation of a regularization parameter, 
we can easily show that the metric on the boundary remains spacelike, $ds'^{2}|_{\CB} >0$ and the corresponding symmetry generators $ \varsigma'^{\pm}_{p}$ preserves the boundary condition 
\begin{align}\label{eq:transBraHannBoundaryCond}
g'_{ab} \sim\begin{pmatrix}
-\frac{R^2}{\ell^2}+ \CO(1) & \CO(r^{-3}) & \frac{R^2}{\ell^2}\Upsilon+\CO(1)\\
& \frac{\ell^2}{R^2} + \CO({r^{-4}})&\CO(r^{-3}) \\
   &  & R^2(1-\frac{\Upsilon^2}{\ell^2})+ \CO(1) \\
\end{pmatrix}.
\end{align}
Performing the asymptotic symmetry analysis with these generators and evaluating the relevant central extension terms on the boundary, we get the same left and right moving central charges as $c_{L,R}={3\ell}/{2G_3}$ for Einstein gravity (and similarly for NMG and TMG). This indicates that the deformation process easily holds true for the asymptotically AdS$_{3}$ black holes ensuring the validity of the procedure adopted in the present work. In fact, it is possible to relate the regularized ASG for asymptotically AdS$_{3}$ case with the one used in \eq{nonchiraldiffeofourierreg}. To see this, we note that locally, NHEBTZ is the same as asymptotic AdS$_{3}$ and the mapping between them is already known (see equation (2.8) of \cite{Balasubramanian:2009bg}). Motivated by this, we consider the following transformations 
\begin{align}\label{T'RPh'tot'rp'}
    T = \frac{\ell}{2}e^{\ph'} + \frac{\ell}{4} 
    \big[t' - \b \ph' + \frac{1}{r} \big], \:
  \Ph = \frac{1}{2}\big[e^{\ph'} - \frac{1}{2} \big( t' - \b \ph' + \frac{1}{r} \big) \big],\:
  R = \ell \sqrt{r}e^{-\ph'/2}.
\end{align}
 This transformation maps the asymptotic AdS$_3$ metric to the regularized  NHEBTZ metric, as described by equation \eq{eq:extremalBTZnearhorreg}.  Furthermore, the action of \eq{T'RPh'tot'rp'} transforms the Brown-Henneaux boundary condition \eq{eq:BraHannBoundaryCond} exactly into the non-chiral boundary condition \eq{nonchiralBC}. This allows us to transform $(\varsigma^{-}_{p}$ and $\varsigma^{+}_{p})$ in $(t',r,\ph')$ coordinates as
\begin{subequations}
\begin{align}
\varsigma^{-}_{p} &= e^{\frac{-ip}{2\b}(t' - \b \ph')} \big[\b \pa_{t'} + ipr\pa_r \big],\label{adstoNHEKright} \\
\varsigma^{+}_{p} &= e^{-\ph'}e^{[-ip\;e^{\ph'}]} \Big[\partial_{\ph'} + \b\pa_{t'}- ip\;r e^{\ph'}\partial_{r} + r \partial_{r}\Big].\label{adstoNHEKleft}
\end{align}
\end{subequations}
The right-hand sides of the above expressions are similar to the one given in \eq{nonchiraldiffeofourierreg}. 
Because of the exponential factor $e^{\ph'}$, the mode function $ \varsigma^{+}_{p}$ needs to be re-expressed in the Fourier series  
    \begin{align}\label{adstoNHEKleftFourier}
     \varsigma^{+}_{p} &= \sum_{q}e^{iq\ph'}\big[ A^{p}_{q}(\partial_{\ph'} +\b\pa_{t'}+r\pa_{r})+ipr B^{p}_{q} \partial_{r} \big],
       \end{align}
       with non-trivial  Fourier coefficients, $A^{p}_{q}$ and $B^{p}_{q}$ given by
        \begin{align}
   A^{p}_{q} \sim \int_{0}^{2\pi} \ d\ph' \ \Big(e^{-[\ph'+ ipe^{\ph'}]}\Big) e^{-iq\ph'} \ ; \   B^{p}_{q} \sim \int_{0}^{2\pi} \ d\ph' \ \Big(e^{-ipe^{\ph'}}\Big) e^{-iq\ph'}.
\end{align}  
Transforming $\varsigma^{+}_{p}$ and   $\varsigma^{-}_{p}$ to $(T',R,\Ph')$ coordinates we can easily establish the local relation between regularized ASG in asymptotic AdS$_{3}$ and the same in NHEBTZ. 
  It is important to note that for asymptotically AdS$_{3}$ black holes both, left as well as right moving sectors are thermally excited, while for NHEBTZ only the left moving part contributes to the Frolov-Thorne temperature.

\section{Conclusions}

In this paper, we have studied the near-horizon symmetries of extremal BTZ black holes in NMG and TMG. We examined the boundary condition put forth in \cite{Azeyanagi:2011zj}. The diffeomorphisms preserving this boundary condition enhance both $U(1)$ as well as $SL(2,R)$ isometries to the Virasoro level, resulting in the non-chiral algebra. However, the associated field representations, or asymptotic symmetry charges, result only in the non-trivial left-moving central charge. This distinctive characteristic arose because the NHEBTZ metric becomes null at the conformal boundary. To address this issue, we considered the most general linear transformation in the $t-\ph$ plane and explicitly showed that the deformed expressions for the BBC potential produce correct central charges, $c_L$, and $c_R$, for NMG and TMG. Moreover, these central charges match exactly with their respective values evaluated by the usual Brown-Henneaux boundary condition at the spatial infinity. Effectively, for both NMG and TMG, we have shown the existence of two centrally extended Virasoro algebras in the near-horizon region of the extremal BTZ black holes. Some remarks regarding our analysis are given below:
\begin{itemize}[leftmargin=*]
\item The boundary condition \eq{nonchiralBC} is appropriate for geometries where $\bar{g}_{tt}$ is either zero or constant. This restricts the choice of coefficients employed in the regularization procedure. Thus, we had just one coefficient, $\b$, which was used as a regularization parameter in our analysis. This aspect is emphasized in the discussion below \eq{genregmetric2}.  
\item The $(t,r)$ component of the conserved 2-form for right-moving ASG was shown to be of order $1/r^3$ for NMG (and $1/r^2$ for TMG), which results in trivial $c_R$. However, the $(\ph,r)$ component remains nonvanishing. As elucidated in the discussion below \eq{eq:kphirforleftrightdiffnonchiral}, the deformation process results in a mixing of $(t,r)$ and $(\ph,r)$ components of the BBC potential. This mixing leads to a nonvanishing central extension for right-moving ASG in the new coordinates while preserving the corresponding expression for left-moving ASG. In short, the delicate balancing of the various components of the BBC potential leads to the emergence of two non-vanishing central charges.  
\item Our successful attempt to provide the non-chiral extension of asymptotic symmetries for extremal black holes in NMG and TMG demonstrates that the higher derivative components in the Lagrangians do not influence the regularization process. 
\item Additionally, the prescription is applicable in scenarios where the conformal boundary is spacelike. This point is discussed in section \ref{4} concerning asymptotically AdS$_{3}$ black holes. In fact, we provided a mapping between the ASG in asymptotically AdS$_{3}$ black holes and the one used in our case. It will be encouraging to see how the different boundary conditions explored already in the literature \cite{Compere:2013bya, Grumiller:2016pqb, Detournay:2023zni} can be related to the present case. 
\end{itemize}
From our study, it appears feasible that the methodology used here may be extrapolated to further higher derivative theories that accommodate the NHEBTZ solution, including Generalized Minimal Massive Gravity \cite{Setare:2016jba} and Extended New Massive Gravity \cite{Sinha:2010ai, Nam:2010dd,Yekta:2024aac}. The null warped black hole examined in \cite{Anninos:2010pm} may signify another prospective topic of research, as this metric aligns with the non-chiral boundary condition employed in this paper. We aim to address these issues in the near future. 
\section*{Acknowledgments}

S.K. is supported by the University Grants Commissions Faculty Recharge Programme (UGC-FRP), Govt. of India, New Delhi, India. This work is supported by the ASPIRE grant of Savitribai Phule Pune University (SPPU), Pune. D.B. is grateful for the hospitality of the Department of Physics, SPPU, Pune, and acknowledges its support. The authors are grateful to Seiji Terashima and Jiaju Zhang for the email correspondence.   

\appendix 
\section{Near horizon limit of extremal BTZ geometry}\label{extremallimit}

In this appendix, we will give explicit calculations of the metric \eq{eq:extremalBTZnearhor}. The near-horizon limits are studied in the literature \cite{Guica:2008mu,Balasubramanian:2009bg,deBoer:2010ac}. 
To get the form of the metric like \eq{eq:extremalBTZnearhor} we need to consider the rotating BTZ black hole whose metric in $(\td{t},\td{r},\td{\ph})$ coordinates represented as 

\begin{align}\label{eq:NonExtremeBTZ}
   ds^2 = - \left(1 - \frac{r_-^2}{\td{r}^2}\right)\left(1 - \frac{r_+^2}{\td{r}^2}\right)\left(\frac{\td{r}^2}{\ell^2}\right) d\td{t}^2 + \left(\frac{\ell^2}{\td{r}^2}\right)\frac{d\td r^2}{\left(1 - \frac{r_+^2}{\td{r}^2}\right)\left(1 - \frac{r_-^2}{\td{r}^2}\right)} + \td{r}^2 \left(d\td{\ph} - \frac{ r_+ r_-}{\ell \td{r}^2} d\td{t} \right)^2,
\end{align}
with $\ell$ as the AdS$_3$ radius and $r_+ \geq r_-$ are the outer and inner horizons of the black hole respectively. 
In the extremal case, we have $r_+=r_-\equiv r_h$, so in that limit the metric becomes 

\begin{align}
   ds^2 = - \left(1 - \frac{r_h^2}{\td{r}^2}\right)^2\left(\frac{\td{r}^2}{\ell^2}\right) d\td{t}^2 + \left(\frac{\ell^2}{\td{r}^2}\right)\frac{d\td r^2}{\left(1 - \frac{r_h^2}{\td{r}^2}\right)^2} + \td{r}^2 \left(d\td{\ph} - \frac{ r_h^2}{\ell \td{r}^2} d\td{t} \right)^2.
\end{align}
Here $r_h$ represents the radius of the event horizon.
Following \cite{deBoer:2010ac} one can take the near-horizon limit as
\begin{align}
    \td{r}^2=r_h^2+\varepsilon \r, \quad \td{t}=\frac{\ta}{\varepsilon} , \quad  \td{\ph}=\varphi+\frac{\ta}{\varepsilon\ell},
\end{align}
and taking $\varepsilon \to 0$ limit gives rise to a spacelike self-dual orbifold. So the resulting metric becomes,
\begin{align}\label{NHEBTZ}
    ds^2=&\frac{\ell^2}{4}\frac{d\r^2}{\r^2}+\frac{\r^2}{r_h^2}\frac{d\ta^2}{\ell^2}+r_h^2\left(d\varphi +\frac{\r}{r_h^2\ell }d\ta \right)^2.
\end{align}
Now redefining the variables as $\ph=({2 r_h}/{\ell})\varphi,\; r={\r}/{r_h^2} \; \text{and}\; t=({2r_h}/{\ell^2})\ta$ we get,

\begin{align}
    ds^2=&\frac{\ell^2}{4}\bigg[-r^2 dt^2+\frac{dr^2}{r^2}+(d\ph-rdt)^2\bigg].
\end{align}
and the $\ph$-direction is compactified as 
\begin{align}\label{phiperodicity}
    (t,\ph)\sim (t,\ph+2\pi \eta),
\end{align}
where $\e$ is a constant given by $\e={2r_h}/{\ell}$ fixed by the mass of the BTZ black hole. This is the form that will be used in the text as the metric for the NHEBTZ black hole. 

 \section{Asymptotic charges}\label{asymptoticcharges}
 In this appendix, we give relevant expressions for the asymptotic symmetry charges for higher derivative gravity such as NMG and TMG.
This can be achieved using the covariant phase space \cite{Iyer:1994ys}, the Barnich-Brandt-Compare (BBC) \cite{Barnich:2001jy, Compere:2018aar} or in the off-shell ADT \cite{Kim:2013zha, Hyun:2014kfa}
formalism. This method is essentially based on equations of motion and the Anderson homotopy operator (see section 1.5 of \cite{Compere:2018aar}). We shall briefly describe this method below.
 
 For a given theory of gravity with action $\CA$ we  define $E^{ab}(g)=\frac{\d \CA}{\d g_{ab}}$ and let
 $\bar{g}_{ab}$ be the background metric satisfying the equations of motion given by $E_{ab}(\bar{g})=0$. Around that background, we expand the full spacetime metric as $g_{ab}=\bar{g}_{ab}+\d g_{ab}$ that in turn leads to order-by-order expansion of $E_{ab}$. From the linear contribution $E^{(1)}_{ab}(g;\bar{g})$ of that expansion  one may construct the asymptotic conserved current corresponding to the asymptotic Killing vector $\xi^b$ as
\begin{align}
    S^a_\xi(\d g;\bar{g})={E^{(1)}}^{ab}(g;\bar{g})\xi_b.
\end{align}
Using $\d g_{ab}= \liexi{\bar{g}_{ab}}=h_{ab}$  the BBC potential $k^{ab}_\xi$ can be extracted as,
\begin{align}\label{surfacechargeshen}
\begin{split}
    k^{ab}_\xi[h;\bar{g}] = &\frac{1}{2}\Ps^A\frac{\pa S^a_\xi}{\pa \Ps^A_b}+\left(\frac{2}{3}\Ps^A_c-\frac{1}{3}\Ps^A\pa_c\right)\frac{\pa S^a_\xi}{\pa \Ps^A_{bc}} + \left(\frac{3}{4} \Ps^A_{cd}-\frac{1}{2}\Ps^A_c\pa_d+\frac{1}{4}\Ps^A\pa_c\pa_d \right)\frac{\pa S^a_\xi}{\pa \Ps^A_{bcd}} \\
    &+\left(\frac{4}{5}\Ps^A_{cde}-\frac{3}{5}\Ps^A_c\pa_d\pa_e+\frac{2}{5}\Ps^A_c\pa_d\pa_e-\frac{1}{5}\Ps^A\pa_c\pa_d\pa_e\right)\frac{\pa S^a_\xi}{\pa \Ps^A_{bcde}}+\hdots-(b\longleftrightarrow a),
\end{split}
\end{align}

where $\Ps^A=h_{mn}$, $\Ps^A_b=\pa_b h_{mn}$, $\Ps^A_{cd}=\pa_c\pa_d h_{mn}$, etc. Note that the factor of $\sqrt{-g}$ is included in the above expression. For TMG, the first three terms contribute, while for NMG the four terms and their antisymmetric counterparts contribute to the BBC potential. It is worth mentioning that for higher derivative or higher curvature theories the conventional Wald-Iyer or BBC 2-form potential needs to be modified appropriately \cite{Azeyanagi:2009wf}. Such a modification is already included in \eq{surfacechargeshen} as it is derived directly from equations of motion. The explicit expression for BBC 2-form derived from the covariant phase space method can be found in \cite{Compere:2008cv} for TMG and in \cite{Detournay:2016gao} in the case of NMG. 

 Integrating $k^{ab}_\xi[h;\bar{g}]$ on codimension-2 surface $\pa\Sigma$ yields the difference between the surface charge 
\begin{align}
    \delta Q_{\xi}[h;\bar{g}] = \int_{\pa\Sigma} \ d\Sigma_{ab} \ 
    k^{ab}_{\xi}[h;\bar{g}] ,
\end{align}
with 
\begin{align*}
    d\Sigma_{ab} = \frac{1}{2!(d-2)!}\;\ten{\epsilon}{}{ab e_1 e_2\hdots e_{d-2}}\;\;dx^{e_1}\wedge dx^{e_2}\wedge \hdots \wedge dx^{e_{d-2}}.
\end{align*}
Assuming the integrability criterion, the total charge $Q_{\xi}$ is obtained by integrating \eq{surfacechargeshen} along a path in the space of metrics satisfying the given boundary conditions.   
\noindent Finally, the central extension $C_{\xi,\zt}$ is calculated using the representation theorem \cite{Compere:2018aar}, as 

 \begin{align}
     [Q_{\xi},Q_{\zt}] &= Q_{[\xi,\zt]} + C_{\xi,\zt}  \ \ \text{with} \ 
     C_{\xi,\zt} = \int_{\pa\Sigma} \ d\Sigma_{ab} \ 
    k^{ab}_{\zt}[\liexi{\bar{g}};\bar{g}].
 \end{align} 
The expression \eq{surfacechargeshen} is especially useful for evaluating the charges and central extension using  Mathematica as shown in \cite{Chen:2013aza, comperemathematica}. 


\end{document}